# Prévision du risque de crédit :
# Une étude comparative entre l'Analyse Discriminante et l'Approche Neuronale

Younés BOUJELBENE, Sihem KHEMAKHEM

Résumé

Les organismes bancaires s'intéressent à évaluer le risque de la détresse financière avant l'octroi d'un crédit. Plusieurs chercheurs ont proposé l'emploi de modèles basés sur les réseaux de neurones en vue d'améliorer la prise de décision du banquier.

L'objectif de cette recherche est d'explorer une nouvelle démarche pratique basée sur les réseaux de neurones en vue d'améliorer la capacité du banquier à prévoir le risque de non remboursement des entreprises demandant un crédit. Cette recherche est motivée par les insuffisances des modèles de prévision traditionnels.

L'échantillon est composé de 86 entreprises tunisiennes et une batterie de 15 ratios financiers a été calculée sur la période 2005-2007. Les prévisions issues de la technique des réseaux de neurones sont comparées à celle de l'analyse discriminante. Les résultats de l'étude montrent que la technique "neuronale" est meilleure en termes de prévisibilité.

**Mots clés:** risque de crédit, prévision, analyse discriminante, réseaux de neurones artificiels.

Abstract

Banks are interested in evaluating the risk of the financial distress before giving out a loan. Many researchers proposed the use of models based on the Neural Networks in order to help the banker better make a decision.

The objective of this paper is to explore a new practical way based on the Neural Networks that would help improve the capacity of the banker to predict the risk class of the companies asking for a loan. This work is motivated by the insufficiency of traditional prevision models. The sample consists of 86 Tunisian firms and 15 financial ratios are calculated, over the period from 2005 to 2007. The results are compared with those of discriminant analysis. They show that the neural networks technique is the best in term of predictability.

# INTRODUCTION

L'évolution de l'activité bancaire durant ces dernières décennies implique l'apparition de divers risques et l'aggravation de ceux déjà existants. Ainsi, la gestion des risques dans le domaine du crédit est un enjeu important, elle est nettement améliorée et contribue à renforcer la solidité financière des établissements de crédit : il s'agit du thème central des nouveaux accords de Bâle II. L'accord de Bâle II est une étape sur la voie menant à l'harmonisation internationale des réglementations bancaires. Il permet entre autre de corriger certaines faiblesses notoires de Bâle I. L'évolution prudentielle du ratio Cooke vers le ratio Mc Donough a pour objectif principal la stabilité financière et la solidité du système bancaire. C'est ainsi que les nouvelles directives de trois piliers fondamentaux de Bâle II, de risque du crédit, de marché et de risque opérationnel incitent l'intensification d'une gestion efficace et saine des risques, tout en tenant compte de l'évolution rapide des marchés financiers. Cependant, dans quelle mesure les nouvelles dispositions pourront réellement assurer la sécurité et la stabilité du système financier international?

Plusieurs types de risques peuvent affecter une banque. Toutefois, le risque de contrepartie, ou risque de crédit, est à la fois le premier, le plus dangereux et le courant risque auquel est confronté un établissement financier. D'une façon générale, le risque de crédit se définit comme étant le risque qu'un emprunteur manque à ses engagements : qu'il soit incapable de tenir sa promesse de verser le paiement des intérêts en temps voulu ou de rembourser le principal à échéance.

De ce fait, et vue que la prise de risque est un synonyme de plus de rentabilité, les banques font une grande partie de leurs profits avec leurs activités de prêts et sont donc très intéressées à développer des modèles d'évaluation du risque de crédit toujours plus précis afin d'optimiser le rendement des prêts consentis.

Plusieurs méthodes ont été proposées pour prévoir le risque de crédit. La technique la plus utilisée est le crédit Scoring à partir de l'analyse discriminante. Cette opération est finalisée par l'engagement d'une fonction score qui aide à la prise de décision dans l'octroi de crédit des entreprises emprunteuses. De nombreuses recherches se sont basées sur l'analyse discriminante (Altman, 1968 ., Altman, Haldeman et Narayanan, 1977 ., Conan et Holder, 1979). Cependant, la méthode de l'analyse discriminante a été critiquée par plusieurs auteurs (Eisenbeis, 1977 ., Deakin, 1976 ., Joy et Tollefson, 1975) parce que la validité des résultats trouvés par cette technique est tributaire de leurs hypothèses restrictives, en l'occurrence l'hypothèse de la normalité de la distribution de chacune des variables retenues et l'hypothèse de l'indépendance entre celles-ci.

Pour pallier aux insuffisances de la méthode de l'analyse discriminante, d'autres modèles d'analyse du risque ont vu le jour. Les réseaux de neurones sont des techniques puissantes de traitement non linéaire de données, qui ont fait leurs preuves dans de nombreux domaines. Le réseau de neurones constitue une nouvelle méthode d'approximation de systèmes complexes, particulièrement utile lorsque ces systèmes sont difficiles à modéliser à l'aide des méthodes statistiques classiques.

La première application des réseaux de neurones à l'estimation du risque de défaillance financière a été réalisée par Bell et Alii (1990). L'utilisation de cette technique s'est ensuite intensifiée par les travaux de Tam (1991) et Altman et al (1994). Plusieurs travaux ont montré que l'approche neuronale offre une meilleure précision prédictive comparée à l'analyse discriminante (Odom et Sharda, 1990 ., Abdou et al, 2008). Cependant, Altman et al (1994) recommandent d'utiliser les deux méthodes (l'approche neuronale et l'analyse discriminante).

Dans ce cadre, ce papier a pour objectif principal de déterminer un modèle de discrimination permettant de détecter la détresse financière des entreprises demandant un crédit auprès des établissements bancaires en utilisant une nouvelle démarche basée sur les réseaux de neurones artificiels et de les comparer à la méthode de l'analyse discriminante afin d'améliorer l'aide à la décision du banquier tunisien.

## I. REVUE DE LA LITTÉRATURE
### I.1. L'analyse discriminante

L'analyse discriminante est une technique statistique qui permet de discriminer entre des observations compte tenu de leurs caractéristiques individuelles. Elle est utilisée afin de classer et/ou prévoir un phénomène et que la variable dépendante est de type qualitative. Son application empirique a commencé depuis les années 1930 avec les travaux de Fisher et Mahalanobis (1936).

L'analyse discriminante consiste à trouver une moyenne pondérée de plusieurs ratios (fonction discriminante), calculée pour chaque entreprise, et qui assure le mieux la distinction entre les entreprises en détresse financière et les entreprises performantes. C'est une méthode utilisée notamment par les banques pour le Scoring.

L'analyse discriminante exige que les données soient indépendantes et normalement distribuées. Par conséquent, sa formule générale est comme suit :

$$Z = \alpha + \beta_1 X_1 + \beta_2 X_2 + \cdots + \beta_n X_n$$

Où : $Z$ représente le score de l'entreprise, $\alpha$ est la constante, $\beta i$ représente les coefficients de la combinaison linéaire des variables explicatives et $Xi$, avec $i = 1 - n$.

Le pionnier de la méthode de crédit Scoring est attribuable à Beaver (1966). Il utilise l'analyse univariée afin de distinguer entre les firmes performantes et les firmes en difficultés jusqu'à 5 ans précédant l'événement de faillite. Celle-ci permet d'affecter les entreprises au groupe des entreprises saines ou à celui en difficultés avec le taux d'erreur le plus faible. Bien que cette méthode fournisse des résultats performants, elle a été énormément critiquée. D'une part, cette approche ne permet pas d'apporter une appréciation globale de la situation de l'entreprise. Le fait de traiter chaque ratio de manière séparée ne permet pas de prendre en considération de manière simultanée l'interdépendance existante entre les différents ratios financiers. D'autre part, la situation financière d'une firme ne peut pas être décrite en totalité à travers un unique ratio quelle que soit l'importance de ce ratio. Malgré toutes ces critiques, cette méthode a été le point de départ pour le développement d'autres modèles tel que le modèle z-score publié par Altman (1968) et qui apparaît le modèle de prédiction des défauts le plus populaire de la littérature. Il calcule une fonction score $Z$ qui est une combinaison linéaire de $n$ ratios financiers, et selon que le score d'une entreprise quelconque serait inférieur ou supérieur à un certain seuil, il affirme si l'entreprise est saine ou en détresse.

Toutefois, Le problème majeur dans l'application de ces méthodes, est que la validité des résultats trouvés par ces techniques est tributaire de leurs hypothèses restrictives qui sont rarement satisfaites dans la vie réelle, en l'occurrence l'hypothèse de la normalité de la distribution de chacune des variables retenues et l'hypothèse de l'indépendance entre celles-ci ce qui peut rendre ces méthodes théoriquement invalides (Huang et al, 2004 ; Šušteršic et al, 2009).

Par conséquent, le caractère contraignant des hypothèses de base nécessaires pour une mise en œuvre efficace de l'analyse discriminante a conduit certains chercheurs à tester l'efficacité d'autres outils statistiques.

### I.2. Les réseaux de neurones artificiels (RNA)

Les RNA sont des outils flexibles et non paramétriques inspirés des systèmes biologiques neuronaux. La naissance du domaine des réseaux de neurones artificiels remonte aux années 1940 avec les travaux de Warren McCulloch et Walter Pitts qui ont montré qu'avec de tels réseaux, on pouvait, en principe, calculer n'importe quelle fonction arithmétique ou logique.

La première application concrète des réseaux de neurones artificiels est survenue vers la fin des années 1950 avec les travaux de Frank Rosenblatt portant sur le perceptron. Un RNA est un outil issu de l'intelligence artificielle, habituellement utilisé en sciences appliquées (biologie, physique, etc.) et qui a fait son entrée en finance au début de la décennie 1990, à coté des méthodes statistiques classiques, en tant que méthode quantitative de prévision.

Les RNA se basent sur l'apprentissage, c'est-à-dire que ces systèmes apprennent par eux-mêmes les relations entre les différentes variables, à partir d'un échantillon de données, en simulant le raisonnement humain. Ils nous permettent de mettre en relation les inputs (la base de données) et les outputs (le résultat) sous la supposition que cette relation est non linéaire. Dans notre cas, la présence de risque de crédit ou non, conformément au schéma ci-dessous :

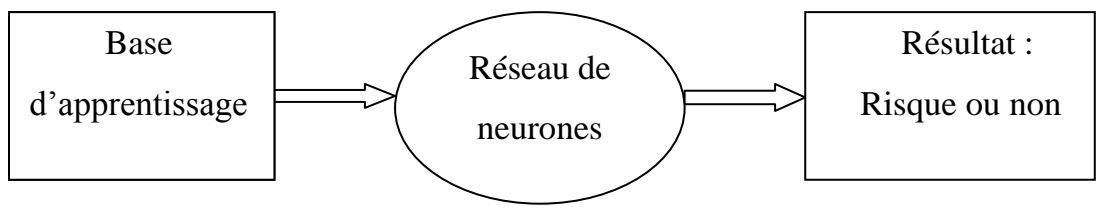

*Figure 1 : Schéma général du traitement*

Un RNA est généralement formé d'une couche d'entrée représentant les neurones d'entrées (variables d'input), d'une couche de sortie représentent le vecteur des variables d'outputs permettant de transférer les informations en dehors du réseau, et d'une ou de plusieurs couches cachées présentant l'ensemble des nœuds cachés ayant des connexions entrantes qui proviennent des neurones d'entrée.

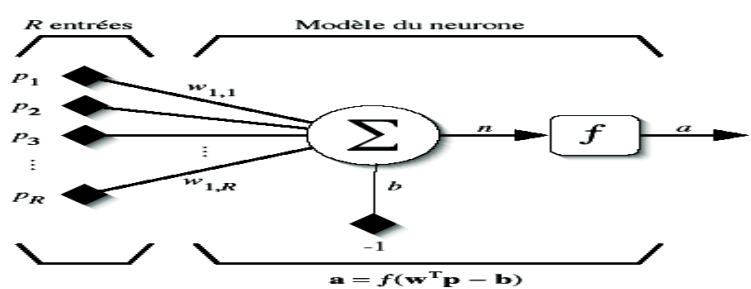

*Figure 2 : Modèle d'un neurone artificiel*

La figure ci-dessus possède $R$ entrées, à chaque entrée $P_i$ est affecté un poids synaptique $w_{1,i}$. Le neurone va commencer par en faire la somme pondérée, c'est sa fonction d'entrée, ce qui va donner son état interne. Le résultat $n$ de cette somme est transformé par une fonction de transfert (appelée aussi fonction d'activation) $f$ qui produit la sortie $a$ du neurone. Cette fonction de transfert est très importante et détermine le fonctionnement du neurone et du réseau. Elle peut prendre différentes formes, ce peut être une fonction à seuil, linéaire ou encore sigmoïdale. On peut considérer un modèle de neurone comme une boîte noire avec des entrées et des sorties. Dans le cas des neurones de McCulloch-Pitts (1943), la fonction de transfert est du type « tout ou rien » (fonction à seuil bipolaire) : si l'activation du neurone est supérieure au seuil, son état final sera +1, si elle est inférieure au seuil, son état final sera -1. Les fonctions à seuil bipolaire ne sont pratiquement plus utilisées. Généralement, on utilise celles de type sigmoïde. Les paramètres (poids des arcs) du RNA ont besoin d'être estimés avant que le modèle ne soit utilisé pour fins de prédiction. Le processus qui consiste à déterminer ces poids est la période d'entraînement ou d'apprentissage. Pour les problèmes de classification, l'apprentissage est dit supervisé dans le sens où les combinaisons d'entrées-sorties désirées sont connues. L'apprentissage supervisé est beaucoup plus rapide que pour les deux autres algorithmes existants (non supervisé ou par renforcement) puisque l'ajustement des poids est fait directement à partir de l'erreur, soit la différence entre la sortie obtenue par le RNA et la sortie désirée ou observée. Un algorithme d'apprentissage supervisé populaire est le processus de rétropropagation des erreurs (back propagation of errors).

Les RNA ont été examinés par Arminger et al (1997), Desai et al (1996), Lee et al (2002), West (2000), Khashman (2010), Tsai et al (2009) et Oreski et al (2012) dans le traitement des problèmes de crédit Scoring. La majorité des études ont montré que les réseaux de neurones sont plus précis, adaptables et robustes que les méthodes statistiques classiques pour l'évaluation du risque de crédit (Oreski et al ,2012).

### I.3. Comparaison entre l'analyse discriminante et l'approche neuronale

La comparaison de ces deux méthodes est présentée à travers les résultats de plusieurs articles. De nombreuses études ont tenté de démontrer l'efficacité de l'une rapport à l'autre.

En partant de l'importance de l'approche neuronale dans la détection des problèmes dans plusieurs domaines, (Paquet, 1997) mentionne qu'il existe deux raisons essentielles qui poussent les chercheurs à s'intéresser à cet outil. Premièrement, cette méthode est plus flexible que certaines méthodes statistiques classiques puisqu' aucune hypothèse n'est nécessaire à propos de la forme fonctionnelle de la relation entre les caractéristiques et la probabilité de défaut ou à propos de la distribution des termes d'erreur et des variables. Deuxièmement, elle représente un instrument adapté pour traiter des problèmes complexes non structurés, d'où l'impossibilité de spécifier, à priori, la forme de la relation entre les variables étudiées.

Angelini et al (2007) voient que l'approche neuronale diffère de la méthode classique de crédit « scoring » principalement dans la nature de la boîte noire et de sa capacité de traiter une relation non linéaire entre les variables. En général, d'après ces auteurs, les réseaux de neurones sont considérés comme une boîte noire à cause de l'impossibilité d'extraire des informations symboliques de leur configuration interne.

Les études d'Odom et Sharda (1990) mentionnent, que les réseaux de neurones sont plus performants que les méthodes statistiques classiques. Le réseau utilisé dans cette étude donne de meilleurs résultats que l'analyse discriminante sur l'échantillon test. En effet, il classe correctement 81,81% des entreprises contre 74,28% pour l'analyse discriminante.

Coats et Fant (1993), dans le cadre de la prédiction de faillites d'entreprises, recommandent d'utiliser aussi bien la méthode d'analyse discriminante et les réseaux de neurones si le problème de l'utilisateur est simplement de réaliser une classification correcte. Par contre, les réseaux de neurones sont plus performants que la méthode d'analyse discriminante si l'on tient compte du type d'erreur commise et du coût associé.

Les résultats obtenus par Altman et al (1994), faisant leur étude sur 1000 entreprises industrielles italiennes entre 1982 et 1992, concluent que l'obtention des meilleurs résultats nécessiterait l'utilisation à la fois des méthodes de réseaux de neurones et la technique de l'analyse discriminante.

Cependant, Swales et Yoon (1992) confirment que l'utilisation simultanée des méthodes de réseaux de neurones et l'analyse discriminante serait assez difficile à mettre dans les tests actuels, puisque ces deux méthodes reposent sur des bases théoriques différentes.

Hoang (2000), dans sa thèse de doctorat, affirme que les méthodes de l'analyse discriminante peuvent accomplir de meilleures performances que les méthodes de réseaux de neurones quand des schémas linéaires sont impliqués dans la tâche de classification, mais que les méthodes par réseaux de neurones sont plus enclines à détecter des schémas non linéaires dans la tâche de classification.

Dans le cadre de la prévision du risque de crédit, Abdou et al (2008) ont procédé à une analyse comparative entre le réseau de neurones et l'analyse discriminante. L'échantillon de l'étude, qui a été fourni par l'une des banques commerciales en Egypte, est constitué de 581 dossiers de crédit de gestion. Les individus de l'échantillon ont été classés en deux groupes : groupe d'entreprises en bonne santé financière (1) et groupe d'entreprises en détresse financière (0). Une batterie initiale formée de 20 variables indépendantes a été sélectionnée par la banque, mais quelques variables avaient des valeurs identiques pour tout l'échantillon d'où elles ont été exclues comme par exemple la durée de l'emprunt qui a été quatre ans dans tous les cas et tous les clients avaient une carte de crédit. De ce fait, les variables sélectionnées ont été réduites à 12. Ainsi, ces variables sont utilisées dans l'analyse discriminante. Le nombre des entreprises bien classées dans ce modèle est égal à 86,75%. A partir de cette batterie de 12 variables, les auteurs ont appliqué la méthode de régression pas à pas (stepwise). Les résultats de cette méthode révèlent que 9 variables seulement présentent un pouvoir de discrimination important. Les variables trouvées sont ensuite utilisées dans l'analyse discriminante. Cette méthode a permis d'avoir un taux de bon classement de l'ordre de 86,92%. Abdou et al (2008) ont conservé les 12 variables pour construire le modèle neuronal. Les résultats de cette approche montrent un taux de bon classement de l'ordre de 93,98% en utilisant le réseau de neurones à 4 nœuds cachés, ainsi qu'un taux de bon classement de l'ordre de 94,84% en utilisant le réseau de neurones à 5 nœuds cachés. Ces auteurs concluent que l'approche neuronale domine la technique de l'analyse discriminante dans la mesure où les réseaux de neurones présentent le pourcentage de bon classement le plus élevé.

## II. MÉTHODOLOGIE : PRÉSENTATION DE L'ÉCHANTILLON ET DES VARIABLES RETENUES

La base de données est composée de 86 entreprises tunisiennes clientes de l'une des banques commerciales tunisiennes à différents secteurs d'activités sur trois ans 2005-2006 et 2007. Afin de tester le pouvoir prédictif de la méthode de discrimination, cet échantillon est subdivisé en deux sous échantillons appelés échantillon de base et échantillon test comportant les informations de 2005- 2006 et 2007 respectivement.

Les entreprises sont classées selon deux groupes en se référant à l'avis du responsable des opérations d'octroi de crédit de la banque: groupe d'entreprises performantes (classe 1) et groupe d'entreprises non performantes (classe 0).

L'échantillon de base comporte 60 entreprises performantes et 26 entreprises non performantes qui sont réparties sur 2005 et 2006 soit un total de 120 observations performantes et 52 observations non performantes. Les données de l'année 2007 comportent 60 entreprises performantes et 26 entreprises non performantes utilisées pour l'échantillon test.

Une batterie initiale formée de 15 ratios, codés de R01 à R15, a été sélectionnée. Ces ratios ont été utilisés par la banque et ont été jugés pertinents dans l'explication de la situation financière des entreprises.

*Tableau 1 : Les variables de l'étude*

|  | **Intitulés des variables** | **Mesure des variables** |
|---|---|---|
| R01 | Taux de valeur ajoutée | Valeur ajoutée/ Chiffre d'affaires |
| R02 | Rentabilité opérationnelle | Excédent brut d'exploitation/chiffre d'affaires |
| R03 | Marge opérationnelle | Résultat d'exploitation / Chiffre d'affaires |
| R04 | Ratio d'endettement | Charges financières/Chiffre d'affaires |
| R05 | Ratio de marge bénéficiaire | Résultat net/ Chiffre d'affaires |
| R06 | Rentabilité financière | Résultat net/Fonds propres nets |
| R07 | Ratio de Solvabilité | Fonds propres nets/ Total bilan |
| R08 | Dépendance financière | Dettes à long et moyen terme/ Capitaux permanents |
| R09 | Capacité de remboursement | Dettes à long et moyen terme/ Cash flow net |
| R10 | Rentabilité économique | Résultat net/Total bilan |
| R11 | Ratio d'immobilisation de l'actif | Actifs immobilisés/ Actif total |
| R12 | Autonomie financière | Capitaux propres/ Capitaux permanents |
| R13 | Marge brute d'autofinancement | Cash flow/ Chiffre d'affaires |
| R14 | Evolution du fonds de roulement | Fonds de roulement/ Chiffre d'affaires |
| R15 | ratio de synthèse | Capitaux permanents/ Immobilisations nets |

Pour construire un score permettant une détection précoce des difficultés de l'entreprise, le choix des variables explicatives est une étape importante qui conditionne l'efficacité de la fonction score. Pour cette raison, nous utilisons une analyse de sélection pas à pas (stepwise) des ratios les plus pertinents permettant de discriminer les deux groupes d'entreprises définies par les valeurs respectives de l'indicateur de difficultés financières.

Les variables retenues à partir de cette sélection et qui seront utilisées dans la suite de ce travail sont les suivantes : R02, R03, R04, R05, R06, R08, R09, R10, R12

### III. RESULTAT DE L'ETUDE

### III.1. Résultat de l'analyse discriminante

Pour vérifier l'existence d'une forte relation entre les ratios utilisés et l'appartenance à un groupe, il faut passer par le test d'égalité des moyennes des groupes. Par ailleurs, les variables les plus discriminantes dans l'analyse doivent avoir des valeurs élevées dans le test de Fisher et, bien évidemment, une significativité qui tendra vers Zéro.

*Tableau 2: test d'égalité des moyennes des groupes*

|     | Lambda de Wilks | F    | ddl1 | ddl2 | Signification |
|-----|-----------------|------|------|------|---------------|
| R02 | 0,992           | 1,36 | 1    | 170  | 0,245         |
| R03 | 0,993           | 1,15 | 1    | 170  | 0,285         |
| R04 | 0,989           | 1,87 | 1    | 170  | 0,174         |
| R05 | 0,979           | 3,64 | 1    | 170  | 0,058         |
| R06 | 0,999           | 0,16 | 1    | 170  | 0,693         |
| R08 | 0,952           | 8,57 | 1    | 170  | 0,004         |
| R09 | 0,984           | 2,7  | 1    | 170  | 0,102         |
| R10 | 0,986           | 2,34 | 1    | 170  | 0,128         |
| R12 | 0,953           | 8,38 | 1    | 170  | 0,004         |

Ce tableau montre la pertinence du ratio R08 « Dettes à long et moyen terme/ Capitaux permanents » ayant un pouvoir discriminant le plus élevé (F de Fisher 8,57), ce qui montre que cette variable influence bien la situation de l'entreprise et permet de différencier entre les deux catégories d'entreprises (performantes et non performantes). L'endettement joue alors un rôle prépondérant dans l'appréciation de la situation financière actuelle et future d'une firme. Ce résultat est confirmé par les études de St-Cyr et Pinsonneault (1997). Ces auteurs avancent que plus une entreprise est endettée, plus il y a de risque qu'elle éprouve des problèmes de solvabilité un jour ou l'autre. D'après ces auteurs, l'utilisation de la dette induit des risques relatifs à la variabilité du rendement et augmente la probabilité d'insolvabilité. Le ratio R12 « Capitaux Propres/Capitaux Permanents » présente aussi un fort pouvoir discriminant. D'après la norme bancaire, ce ratio doit être supérieur ou égal à 0,5. Dans le cas contraire, on dit que la capacité d'endettement à long terme de l'entreprise est saturée. En termes financiers, les entreprises en détresse souffrent ainsi d'un poids élevé de l'endettement à court terme et d'une faible autonomie financière (Lelogeais, 2003). D'après les résultats, on remarque que certaines variables possèdent un très faible pouvoir discriminant et un taux de signification d'erreur dépasse 0,05. En effet, le ratio R06 « Résultat net/Fonds propres nets » a le pouvoir discriminant le plus faible. Une étude récente menée par Vernimmen (2002) confirme de tel résultat. Selon cet auteur, ce critère bien que séduisant sur le plan de la simplicité de calcul, n'est pas exempt de toute reproche. En effet, et par opposition à d'autres indicateurs, il ne prend pas compte le risque. Il est souvent limité à un seul exercice. De plus, il doit être comparé avec des taux exigés pour être significatif. Le ratio R02 « Excédent brut d'exploitation/chiffre d'affaires» a un pouvoir discriminant faible. Cependant, Stili (2002) mentionne que l'excédent brut d'exploitation est le premier solde obtenu à l'issu du processus de production et de commercialisation. Ce concept est la principale composante pour le calcul du ratio de la rentabilité opérationnelle. Celui-ci constitue une première mesure directe de la performance industrielle et commerciale de l'entreprise et un indicateur, généralement significatif, de sa capacité bénéficiaire.

Pour établir l'équation discriminante et déterminer un score, on a estimé les coefficients de la fonction discriminante à partir du tableau 3.

*Tableau 3 : coefficients des fonctions discriminantes canoniques*

|  | Fonction 1 |
|---|---|
| R02 | 1,671 |
| R03 | -0,779 |
| R04 | -0,566 |
| R05 | -6,151 |
| R06 | 0,087 |
| R08 | 1,364 |
| R09 | 0,008 |
| R10 | 0,005 |
| R12 | 0,037 |
| (Constante) | -0,188 |

Cela permet d'établir la fonction discriminante suivante :

$$D1\,(i0) = -0{,}188 + 1{,}671 R02 - 0{,}779 R03 - 0{,}566 R04 - 6{,}151 R05 + 0{,}087 R06 + 1{,}364 R08 + 0{,}008 R09 + 0{,}005 R10 + 0{,}037 R12$$

Une des mesures de pouvoir discriminant est le taux de bon classement. Ce dernier est égal au rapport du nombre d'entreprises bien classées dans les deux groupes sur le nombre total d'entreprises.

*Tableau 4 : résultats de classement de l'échantillon de base*

| Y | | | Classe(s) d'affectation prévue(s) | | Total |
|---|---|---|---|---|---|
| | | | 0 | 1 | |
| Original | Effectif | 0 | 10 | 42 | 52 |
| | | 1 | 2 | 118 | 120 |
| | % | 0 | 19,231 | 80,769 | 100 |
| | | 1 | 1,667 | 98,333 | 100 |

a 74,4% des observations originales classées correctement.

Comme le montre ce tableau, le taux de bon classement des entreprises non performantes est égal à 19,231% et le taux de bon classement des entreprises performantes est égal à 98,33%. Ainsi, le nombre d'entreprises bien classées par le modèle est égal à 74,4%.

### III.2. Résultat des réseaux de neurones artificiels

Pour créer, manipuler et visualiser des résultats obtenus par les réseaux de neurones, nous avons utilisé le logiciel *Matlab 7.1*. Ce dernier comprend une application *"neural network toolbox"*, qui permet la modélisation des réseaux de neurones artificiels.

L'architecture du réseau de neurones de type perceptron multicouches est utilisée pour construire les modèles de prévision de la détresse financière. Pour pouvoir déterminer la meilleure architecture, nous avons utilisé la fonction « trainrp » comme fonction d'apprentissage. En effet l'algorithme choisi est l'algorithme de rétropropagation du gradient. Ce dernier est entraîné sur tout l'ensemble d'apprentissage. Par ailleurs, la fonction d'activation que nous avons choisi pour notre application est la fonction sigmoïde « logsig » pour les neurones cachés dans les couches cachées et la fonction linéaire « purelin » pour le neurone de sortie. La fonction de création d'un réseau est « newff », pour feedforward où les neurones d'inputs ne sont pas connectés avec ceux d'outputs. Cette commande crée le réseau et initialise ses poids. Enfin, nous avons retenu comme fonction de performance la moyenne des erreurs quadratiques (MSE) :

$$E = \frac{1}{2}\sum_{i=1}^{n}(d_i - y_i)^2$$

Avec, $d_i$: valeur désirée de l'output, $y_i$: valeur calculée de l'output et $n$: nombre d'observations dans l'échantillon

Cette équation détermine l'erreur quadratique moyenne commise par le modèle neuronal. C'est à partir de cet indicateur que le réseau décide de continuer ou non à rechercher la solution désirée. L'idéal c'est d'avoir une erreur très faible voir même nulle. La fonction « sim » est utilisée pour calculer la MSE de l'échantillon test. On va prendre comme input de la fonction d'apprentissage les 9 variables dégagées pertinentes et une variable de sortie qui prendra une valeur 1 ou 0 selon que l'entreprise est considérée comme saine ou en difficulté, au cours du processus d'apprentissage.

De ce fait, dans ce travail, on a essayé de faire plusieurs tests sur le réseau en faisant varier le nombre de couches cachées et le nombre de neurones cachés dans chaque couche afin de choisir la meilleure architecture qui présente un taux d'erreur minimal puisqu'il n'existe aucune loi, aucune règle, aucun théorème qui permettrait de déterminer le nombre de couches cachées et le nombre de neurones à placer dans la couche cachée pour avoir un réseau de neurones optimal. En effet, dans notre programme, nous avons fixé un nombre d'itération égal à 500 et un nombre de couches cachées minimal égal à 1 jusqu'au 5. Le tableau suivant récapitule les résultats auxquels nous avons abouti.

*Tableau 5 : Récapitulation des résultats des réseaux de neurones*

| Réseaux multicouches : feedforward architecture | Nombre de couches totales | Nombre de couches cachées | MSE de l'échantillon d'apprentissage[1] | MSE de l'échantillon test[2] |
|---|---|---|---|---|
| Net1_1 [9 1 1] | 3 | 1 | 0,1485 | 0,16038 |
| Net1_2 [9 3 1] | 3 | 1 | 0,1249 | 0,14437 |
| Net1_3 [9 4 1] | 3 | 1 | 0,1013 | 0,1053 |
| Net1_4 [9 6 1] | 3 | 1 | 0,1046 | 0,11895 |
| Net1_5 [9 7 1] | 3 | 1 | 0,0807 | 0,1257 |
| Net1_6 [9 4 6 1] | 4 | 2 | 0,0569 | 0,09744 |
| **Net1_7 [9 6 8 1]** | **4** | **2** | **0,0086** | **0,0608** |
| Net1_8 [9 2 4 5 1] | 5 | 3 | 0,1298 | 0,14756 |
| Net1_9 [9 5 6 7 1] | 5 | 3 | 0,0671 | 0,08604 |
| Net1_10 [9 2 3 4 3 1] | 6 | 4 | 0,1384 | 0,1523 |
| Net1_11 [9 3 4 4 4 1] | 6 | 4 | 0,0846 | 0,11891 |
| Net1_12 [9 1 2 3 4 1 1] | 7 | 5 | 0,1106 | 0,16454 |

D'après ce tableau, nous remarquons que le nombre de couches optimal est de 4 dont 2 intermédiaires. Ce réseau nous a permis d'avoir une erreur moyenne quadratique le moins élevée aussi bien pour l'échantillon d'apprentissage et celui de l'échantillon test qui s'élèvent respectivement à 0,00868 et 0,0608 et un taux de bon classement de l'ordre de 80,23%.

Dans ce travail, le pourcentage de bon classement de l'échantillon test de la meilleure architecture est déterminé en calculant la médiane pour la valeur $y_i$ pour faire le regroupement des entreprises.

D'où, chaque valeur de $y_i$ supérieur à la médiane prend la valeur 1 et chaque valeur de $y_i$ inférieur à la médiane prend la valeur 0 :

- Si $y_i \geq 0,785$ : l'entreprise est considérée saine
- Si $y_i \leq 0,785$ : l'entreprise est considérée en difficulté

---

[1] Erreur quadratique moyenne commise lors du classement des entreprises de l'échantillon d'apprentissage

[2] Erreur quadratique moyenne commise lors du classement des entreprises de l'échantillon test

*Tableau 6 : Résultat de classement de la meilleure architecture*

|  | 1 | 2 | 3 | 4 | 5 | 6 | 7 | 8 | 9 | 10 | 11 | 12 |
|---|---|---|---|---|---|---|---|---|---|---|---|---|
| $d_i$ désirée | 1 | 1 | 1 | 1 | 1 | 1 | 1 | 1 | 1 | 1 | 1 | 1 |
| $y_i$ réel | 0,852 | 0,785 | 0,798 | 0,795 | 0,791 | 0,85 | 0,808 | 0,896 | 0,799 | 0,008 | 0,892 | 0,865 |
|  | 1 | 1 | 1 | 1 | 1 | 1 | 1 | 1 | 1 | 0 | 1 | 1 |
|  | 13 | 14 | 15 | 16 | 17 | 18 | 19 | 20 | 21 | 22 | 23 | 24 |
| $d_i$ désirée | 1 | 1 | 1 | 1 | 1 | 1 | 1 | 1 | 1 | 1 | 1 | 1 |
| $y_i$ réel | 0,516 | 0,873 | 0,848 | 0,008 | 0,785 | 0,882 | 0,818 | 0,803 | 0,785 | 0,791 | 0,785 | 0,852 |
|  | 0 | 1 | 1 | 0 | 1 | 1 | 1 | 1 | 1 | 1 | 1 | 1 |
|  | 25 | 26 | 27 | 28 | 29 | 30 | 31 | 32 | 33 | 34 | 35 | 36 |
| $d_i$ désirée | 1 | 1 | 1 | 1 | 1 | 1 | 1 | 1 | 1 | 1 | 1 | 1 |
| $y_i$ réel | 0,884 | 0,809 | 0,832 | 0,828 | 0,785 | 0,743 | 0,304 | 0,864 | 0,873 | 0,887 | 0,785 | 0,876 |
|  | 1 | 1 | 1 | 1 | 1 | 0 | 0 | 1 | 1 | 1 | 1 | 1 |
|  | 37 | 38 | 39 | 40 | 41 | 42 | 43 | 44 | 45 | 46 | 47 | 48 |
| $d_i$ désirée | 1 | 1 | 1 | 1 | 1 | 1 | 1 | 1 | 1 | 1 | 1 | 1 |
| $y_i$ réel | 0,262 | 0,151 | 0,899 | 0,039 | 0,872 | 0,191 | 0,817 | 0,617 | 0,785 | 0,859 | 0,869 | 0,54 |
|  | 0 | 0 | 1 | 0 | 1 | 0 | 1 | 0 | 1 | 1 | 1 | 0 |
|  | 49 | 50 | 51 | 52 | 53 | 54 | 55 | 56 | 57 | 58 | 59 | 60 |
| $d_i$ désirée | 1 | 1 | 1 | 1 | 1 | 1 | 1 | 1 | 1 | 1 | 1 | 1 |
| $y_i$ réel | 0,833 | 0,926 | 0,95 | 0,835 | 0,85 | 0,877 | 0,785 | 0,563 | 0,862 | 0,897 | 0,868 | 0,041 |
|  | 1 | 1 | 1 | 1 | 1 | 1 | 1 | 0 | 1 | 1 | 1 | 0 |
|  | 61 | 62 | 63 | 64 | 65 | 66 | 67 | 68 | 69 | 70 | 71 | 72 |
| $d_i$ désirée | 0 | 0 | 0 | 0 | 0 | 0 | 0 | 0 | 0 | 0 | 0 | 0 |
| $y_i$ réel | 0,523 | 0,333 | 0,325 | 0,466 | 0,788 | 0,072 | 0,27 | 0,069 | 0,433 | 0,192 | 0,366 | 0,371 |
|  | 0 | 0 | 0 | 0 | 1 | 0 | 0 | 0 | 0 | 0 | 0 | 0 |
|  | 73 | 74 | 75 | 76 | 77 | 78 | 79 | 80 | 81 | 82 | 83 | 84 |
| $d_i$ désirée | 0 | 0 | 0 | 0 | 0 | 0 | 0 | 0 | 0 | 0 | 0 | 0 |
| $y_i$ réel | 0,162 | 0,111 | 0,132 | 0,878 | 0,287 | 0,008 | 0,558 | 0,441 | 0,445 | 0,812 | 0,821 | 0,611 |
|  | 0 | 0 | 0 | 1 | 0 | 0 | 0 | 0 | 0 | 1 | 1 | 0 |
|  | 85 | 86 |  |  |  |  |  |  |  |  |  |  |
| $d_i$ désirée | 0 | 0 |  |  |  |  |  |  |  |  |  |  |
| $y_i$ réel | 0,008 | 0,24 |  |  |  |  |  |  |  |  |  |  |
|  | 0 | 0 |  |  |  |  |  |  |  |  |  |  |

▨ *Entreprise mal classée*

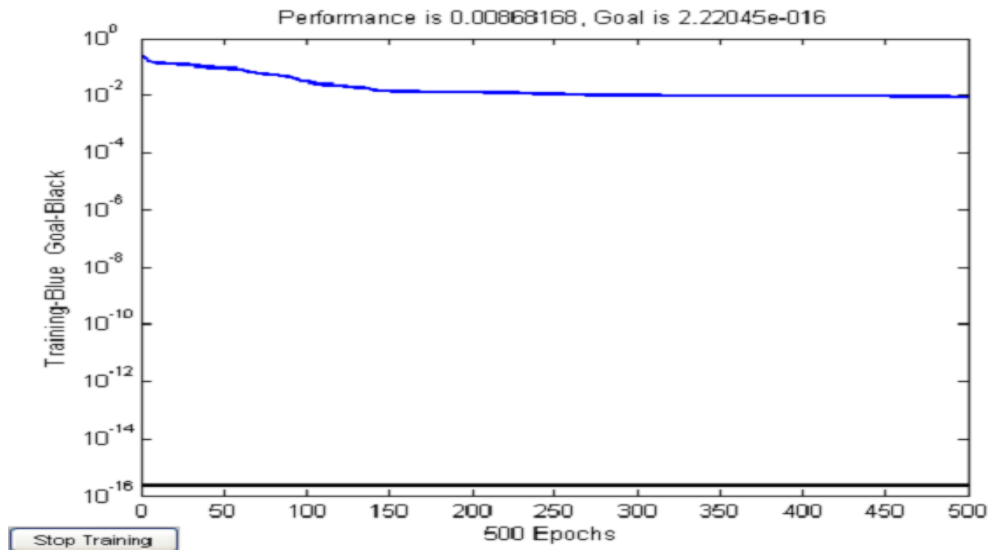

*Figure 1 : Courbe d'apprentissage du réseau optimal à deux couches cachées et à sortie unique*

**III.3. Comparaison de la performance des modèles**

La comparaison des deux modèles (analyse discriminante et l'approche neuronale) en terme de prévisibilité montre la performance de la technique neuronale par rapport à l'analyse discriminante. En effet, le pourcentage de bon classement, issu de l'application des réseaux de neurones artificiels, est meilleur que l'analyse discriminante.

| *Résultat de l'analyse discriminante* | *Résultat des réseaux de neurones* |
|---|---|
| 74,4% | 80,23% |

Ainsi, les réseaux de neurones artificiels apparaissent comme étant un outil de prévision puissant en matière de détresse financière des firmes. Ce travail confirme ainsi les études empiriques déjà établies (Kerling et Podding, 1994 ., Oden et Sharada, 1990 ., Abdou et al, 2008).

Les modèles des réseaux de neurones artificiels sont de plus en plus utilisés en scoring avec des succès divers. D'après certains statisticiens, si ces nouvelles méthodes sont intéressantes et parfois plus performantes que les techniques statistiques traditionnelles, elles sont aussi moins robustes et moins bien fondées. De plus, les réseaux de neurones sont incapables d'expliquer les résultats qu'ils fournissent. Enfin, ils se présentent comme des boîtes noires dont les règles de fonctionnement sont inconnues. Ils créent eux-mêmes leur représentation lors de l'apprentissage. En termes d'interprétation des pondérations, l'analyse discriminante semble être plus performante. En effet, dans un réseau de neurones artificiels, les liaisons internes n'ont pas de signification économique. Les pondérations des ratios figurant dans les fonctions discriminantes sont par contre transparentes et faciles à interpréter du point de vue de l'analyste financier.

Pour conclure, on peut dire que l'approche neuronale et l'analyse discriminante se révèlent être deux techniques complémentaires. L'analyse discriminante nous permet de sélectionner les variables les plus pertinentes et le réseau de neurones peut reprendre ses variables et calcule un taux d'erreur le moins élevé.

**CONCLUSION**

A ce niveau, en peut affirmer que l'approche neuronale est plus performante que l'analyse discriminante en terme de prévision du risque de crédit.

A la lumière des résultats de cette étude, nous pouvons émettre les grandes critiques suivantes:

- Le présent travail peut être étendu en tenant compte d'un plus grand nombre et d'une plus grande variété de variables, notamment, celles qualitatives. Bauer et al (1998) notent que l'analyse par les ratios financiers n'est pas une science exacte dans la mesure où les priorités en terme de calcul varient d'un examen à l'autre. Ces auteurs font référence au recours aux facteurs quantitatifs en tant que guide quant à l'orientation des institutions face au risque de crédit, orientation conservant sans équivoque une part de son caractère imprévisible.

- L'appréciation de la banque en matière d'octroi de crédit est biaisée. En premier lieu, les ratios utilisés par la banque pour évaluer les entreprises ne donnent pas une vue globale de la situation financière puisqu'il existe d'autres variables qui ne sont pas exploitées par la banque malgré qu'elles puissent refléter la réalité de la situation financière des entreprises. En deuxième lieu, la banque accorde dans la pratique des crédits sans aucune référence théorique ou scientifique solide. Parfois, il y a des dossiers de crédits qui sont acceptés malgré qu'ils présentent des difficultés financières que ce soit parce que le client possède des relations avec le personnel de la banque ou bien parce qu'il s'agit d'un client dépositaire ce qui mène à des problèmes d'insolvabilité. Enfin, la banque évalue sa clientèle seulement à travers les critères de sa situation financière. Cependant, Coats et Fant (1993) ont pris l'avis des auditeurs sur la situation financière des entreprises comme critère pour identifier les entreprises en détresse. Mais, cet avis peut conduire à des résultats fallacieux, puisque d'après Altman et al (1994), ce critère est très subjectif et sujet à équivoque.

Plusieurs extensions à l'analyse discriminante et aux réseaux de neurones artificiels sont envisagées. Elles sont de nature à améliorer la prévision et à pallier les inconvénients de ces derniers. Il s'agit, notamment des algorithmes génétiques et des séparateurs à vastes marges qui sont aussi appliqués à la prévision de la détresse financière des firmes.

## Bibliographie